\documentstyle[aps,epsfig]{revtex}

\newcommand{\dd}{\text{d}}

\newcommand{\ee}{\text{e}}

\newcommand{\p}{\partial}

\newcommand{\eps}{\varepsilon}

\begin{document}

\newenvironment{dessins}{{\bf Figures}}{}
\begin{center}
{\large{\bf On the velocity distributions of the one-dimensional inelastic gas}}
\end{center}
\begin{center}
{A. Barrat $^{(a)}$, T. Biben$^{(b)}$, 
Z. R\'acz$^{(a,c)}$, E. Trizac$^{(a)}$ and 
F. van Wijland$^{(a)}$}
\end{center}

\noindent {\small $^{(a)}$Laboratoire de Physique Th\'eorique, Universit\'e de 
Paris-Sud, 91405 Orsay cedex, France.

\noindent $^{(b)}$Laboratoire de Spectrom\'etrie Physique, B\^atiment 45, Avenue de la Physique,
Domaine Universitaire, \\
BP 87, 38402 Saint Martin d'Heres, France.

\noindent $^{(c)}$Institute for Theoretical Physics, E\"otv\"os University, P\'azm\'any
s\'et\'any 1/a, 1117 Budapest, Hungary.}

\vskip 1 mm
\begin{center}{\bf Abstract}\\
\end{center}
{\small 
We consider the single-particle velocity distribution of a
one-dimensional fluid of inelastic particles.
Both the freely evolving (cooling) system and the
non-equilibrium stationary state obtained in the presence of
random forcing are investigated, and special emphasis is paid to
the small inelasticity limit. The results are obtained from
analytical arguments applied to the Boltzmann equation
along with
three complementary numerical techniques (Molecular Dynamics,
Direct Monte Carlo Simulation Methods and iterative solutions of
integro-differential kinetic equations).
For the freely cooling fluid, we investigate in detail
the scaling properties of the bimodal velocity distribution emerging
close to elasticity and calculate the scaling function associated
with the distribution function. In the heated steady state, we find
that, depending on the inelasticity, the distribution function
may display two different stretched exponential tails at
large velocities. The inelasticity dependence of the crossover
velocity is determined and it is found that the extremely high velocity
tail may not be observable at ``experimentally relevant'' inelasticities.}

\vskip 5 mm


\section{Introduction}
\subsection{Motivation}

In the widely studied context of nonequilibrium stationary
states, granular gases stand out as an interesting model system,
accessible to and subject of many experimental and analytic
investigations. Their theoretical description and understanding is
one of the presently important issues of the development of the
out-of-equilibrium statistical mechanics. 

The main difference between molecular gases and granular gases
stems from the fact that at each collision
between e.g. steel or glass beads (in experiments), or idealized
smooth hard spheres (in analytical and numerical investigations), a fraction
of the relative kinetic energy is lost \cite{Jaeger}. 
This inelasticity is responsible
for many interesting phenomena, such as the appearance of spatial
heterogeneities, or non-Gaussian velocity distributions...
Theoretically, two opposite situations have been extensively studied
in the context of smooth inelastic hard spheres we shall consider here.
Namely, the free cooling case where no forcing mechanism compensates
the energy loss due to dissipative collisions (see e.g. the review 
\cite{Dufty} and references therein), 
and the uniformly heated system where an external random force 
acts as a heating process on the grains, allowing a non-equilibrium 
stationary state to be reached \cite{mac,vNE,Peng}.

In this work, we will study the two above situations (i.e. with or 
without energy input), and 
concentrate on a one-dimensional granular fluid. 
For the homogeneously heated gas (section \ref{sec:heated}),
the focus will be on the high energy tail of the velocity distribution 
$P(v)$. Whereas the velocities up to the thermal scale obey 
a Maxwell-Boltzmann-like distribution, we will show 
combining kinetic theory arguments and numerical simulations
(both Molecular Dynamics and Monte Carlo) 
that in the limit of vanishing
inelasticity, $P(v)$ displays a $\exp(-v^3)$ large $v$ behaviour.
At finite inelasticity, this tail is asymptotically 
dominated by the law $\exp(-v^{3/2})$ already predicted in \cite{vNE}.
These predictions will also be confirmed by the results of a high precision 
iterative solution of the non-linear Boltzmann equation. 
On the other hand, without energy injection (section \ref{sec:free}), 
we will similarly
concentrate on the limit of small inelasticity (that appears quite singular,
unlike in the heated case),
and shed some light 
on the importance of spatial heterogeneities and velocity 
correlations: detailed scaling properties 
of the solutions of the homogeneous Boltzmann equation will be obtained 
analytically and checked numerically. Further confrontation against
Molecular Dynamics simulations will show that the velocity distributions of the
Boltzmann homogeneous cooling state share some common features with
those obtained by integrating the exact equations of motion. 

\subsection{The model}

We shall consider a one-dimensional gas of  
equal mass particles of length $\sigma$ and density $n$, evolving on a
line of length $L=N/n$ with periodic boundary conditions.
These particles undergo binary collisions with conservation of momentum
but loss of a fraction ($\alpha^2$) of the kinetic energy 
in the center of mass frame: consequently, if 
$v_1$ and $v_2$ (resp. $v_1^\prime$ and $v_2^\prime$) are the velocities of
the two particles involved before (resp. after) the collision,
then
\begin{eqnarray}
v_1^\prime+v_2^\prime &=&  v_1+v_2\\
v_1^\prime-v_2^\prime &=&-\alpha \,(\,v_1-v_2\,)\,,
\label{rules}
\end{eqnarray}
where $0\leq\alpha\leq 1$ is the restitution coefficient. We also
introduce  the inelasticity parameter $\eps=1-\alpha$ ($\eps=0$ for elastic
collisions).

We will focus on the behaviour of  the 
velocity distribution $P(v, t)$ in two cases:

\begin{itemize}
\item without energy injected, the above collision rules define 
a system where energy 
dissipation through collisions is not balanced and the typical 
velocities of particles progressively decrease. This free cooling regime
has been widely studied 
\cite{Mcnamara1,Mcnamara2,Goldhirsch,Sela,Du,Ramirez,Caglioti,Benedetto,brey1,esipov,%
brey2,montanero,huthmann-orza-brito,bennaim}
and in particular in dimension 1 by molecular dynamics studies 
\cite{Mcnamara1,Mcnamara2,Sela,bennaim}.
Slight modifications of the collision rule allow to bypass the inelastic 
collapse \cite{Bernu} and to observe an asymptotic scaling 
regime for $P(v, t)$ \cite{bennaim}.

\item a steady state can be reached if the loss of energy 
through collisions is balanced by an injection that can be
achieved through a random force $\eta(t)$ acting 
on each particle
\begin{equation}
\frac{\dd v}{\dd t}= F + \eta(t),\;\;\;\langle \eta(t)\eta(t')\rangle 
=2D\delta(t-t')
\label{randomforces}
\end{equation}
where $D$ is the amplitude of the injected power
and $F$ the systematic force due to inelastic collisions.
Velocities consequently execute a random walk in-between the 
collisions and in
the collisionless case $P(v,t)$ obeys a diffusion equation with  
a ``diffusion'' coefficient $D$.
This model was first introduced and discussed by Williams and
MacKintosh~\cite{mac} in dimension 1, and 
studied in higher dimensions \cite{Peng}; 
variants have also been proposed
\cite{puglisi,Cafiero}.
\end{itemize}

\noindent We define the granular temperature as the average kinetic 
energy of the system:
\begin{equation}
T(t) = \int \dd v v^2 P(v,t) \ .
\end{equation}
The function $T(t)$ decreases for the freely cooling system,
but it eventually fluctuates around a steady-state value in the 
heated case.

\subsection{Investigation methods}

Our study relies on three complementary approaches: 

\begin{itemize}
\item Molecular Dynamics (MD) simulations \cite{Allen}
integrate the exact equation of motion in a finite box: we consider
$N$ hard rods of length $\sigma$, on a line of linear size $L$,
with periodic boundary conditions, random initial
velocities, and we use an event-driven algorithm to study their dynamics.

\item the Boltzmann equation describes the evolution of the one-particle
distribution function $P(v,t)$, upon the
molecular chaos factorization hypothesis \cite{Resibois}. This equation is therefore
a mean-field approximation of the problem. It can be solved
numerically by the Direct Simulation Monte Carlo (DSMC) method \cite{Bird},
or in certain cases with an even better precision by an iterative
method similar to that used in \cite{Biben}. 

\item in the elastic limit $\alpha \to 1$, an analytical scaling
approach can be used to study
the Boltzmann equation. 
It is important to note that, in the particular case of one-dimensional
hard spheres, the elastic case $\eps = 0$ is quite peculiar. 
Indeed, for $\eps = 0$, the
collisions only exchange the velocities of the particles: this system
is therefore unable to thermalize and is
equivalent to one with transparent particles where $P(v,t)$ is frozen in time.
\end{itemize}

\section{Steady state of the heated fluid}
\label{sec:heated}

The exact solution of the problem of a $d=1$ inelastic gas appears 
inaccessible, which prompted us to carry out
molecular dynamics (MD) 
simulations to calculate $P(v,t)$. In order to see whether a mean-field 
type approach can give a reasonable description of the inelastic gas,
we reconsider the kinetic theory of the process.
We solve 
the appropriate Boltzmann equation by simulations in the general 
$\alpha$ case, and derive exact results in the $\eps\rightarrow 0$ limit.

Since the gas systematically reaches a stationary state with a temperature $T$
that depends on the inelasticity (see below), it is useful to introduce the
rescaled velocity
\begin{equation}
c = \frac{v}{\sqrt{T}}
\end{equation}
and the corresponding distribution function 
\begin{equation}
f(c)= \sqrt{T(t)} P(v,t).
\end{equation}
In order to compare the velocity distributions at various inelasticities,
numerical data for the {\em rescaled velocity} distributions
will always be displayed in the figures.

\subsection{General $\alpha$ case} 

\subsubsection{Molecular Dynamics simulations}

The molecular dynamics simulations are carried out with $N=5000$ hard rods,
using an event-driven algorithm, and submitting the rods to random
kicks at a frequency that remains much higher than the collision frequency,
in order to simulate the noise of equation (\ref{randomforces}). 
Note that, since the Langevin equation 
considered in \cite{puglisi} is different from ours \footnote{It contains
a friction term in velocity space whose amplitude is linked to the force
amplitude, whereas detailed balance is not satisfied by our model, where the
forcing is independent of the state of the system}, 
comparing the two approaches is not possible without further
investigations on either side.

\begin{figure}[ht]
\centerline{ 
\epsfig{figure=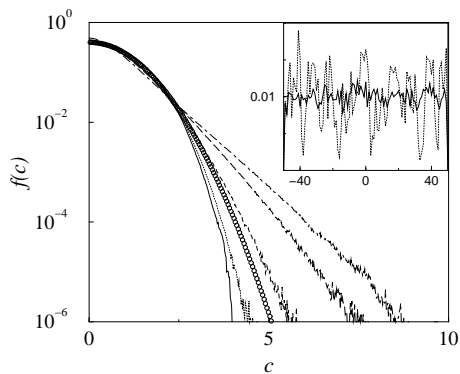,width=6cm,angle=0}    }
\vskip 2mm       
\caption{Velocities distributions for MD simulations with 5000 particles,
density $n \sigma=0.5$,
for restitution coefficients $0.6$, $0.9$, $0.95$,
$0.99$, $0.999$ (from top to bottom). The symbols
show the Gaussian distribution. In the inset are shown the space density
fluctuations for $0.99$ (continuous line) and $0.6$ (dotted line).}
\label{fig:heatedmd}
\end{figure}

Starting from an initial distribution of velocities, a steady-state is
reached after a transient, and velocity distributions can be measured.
Figure \ref{fig:heatedmd} displays such distributions for inelasticity ranging
from $\alpha=0.6$ up to $0.999$. Strong non-Gaussian behaviour is observed, 
with
over- or under-populated high energy tails depending on $\alpha$. Moreover, 
the inset
shows that the system remains quite homogeneous at low inelasticity, while
strong density fluctuations develop at larger inelasticity.
The value of the density $n$ of the system seems to have no influence on the 
shape of the velocity distributions, density fluctuations seem to increase 
roughly proportional to $1/(n\sigma)$ (at constant $\alpha$).
Detailed investigations of the spatial correlations are left for future
studies.

\subsubsection{Kinetic theory}

Assuming that the density of the particles is low and 
neglecting both velocity and spatial correlations of colliding partners,
the following Boltzmann equation for the spatially 
averaged velocity distribution function $P(v,t)$ can be written as
\cite{Mcnamara2,esipov}
\begin{equation}\label{ludwig}
\p_t P-D\p_v^2P=
-n\int\limits_{-\infty}^\infty\dd v'|v-v'|P(v)P(v')
+\frac{4n}{(1+\alpha)^2}\int\limits_{-\infty}^\infty\dd
v'|v-v'|P(v')P\left(\frac{2v-(1-\alpha)v'}{1+\alpha}\right).
\end{equation}
The right-hand side above contains the collision terms corresponding to the 
``dissipative'' rules (\ref{rules}), while the Fokker-Planck term
$D\p_v^2P$ on the left-hand side takes into account the
energy injected by the random forces (\ref{randomforces}).

The system described by Eq. (\ref{ludwig}) is expected to relax 
to a steady state since the power input is independent of the velocities
while the loss of energy is roughly proportional to the energy 
itself. This expectation 
can be made more quantitative by deducing the equation for the 
temperature ($T=\langle v^2 \rangle$) of the system
\begin{equation}\label{exactTemp}
\frac{\dd\langle v^2\rangle}{\dd t}=2 D-\frac{n}{4}(1-\alpha^2)\langle
|v-v'|^3\rangle \, 
\end{equation}
where $\langle...\rangle$ denotes averaging with $P(v,t)$, and $|v-v'|$ 
represents the relative velocity of two randomly chosen 
particles. There is a stationary solution to this equation that 
has a simple physical meaning. Namely the rate of input of energy
$(\sim D)$ is equal to the rate of loss of kinetic energy in the 
center-of-mass frame $\sim (1-\alpha^2)
(v-v')^2$ (with the extra factor $n|v-v'|$ coming from the collision rate). 
One may also estimate 
the characteristic time of reaching the steady state. Indeed,
the quantities $T^{3/2}=\langle v^2\rangle^{3/2}$ and 
$\langle|v|^3\rangle$
are expected to have the same leading large-time ($t\rightarrow\infty$) 
dependence, and thus, up to an unknown numerical constant ${\cal C}$, 
Eq. (\ref{exactTemp}) can be written as
\begin{equation}
\frac{\dd T}{\dd t}=2D-{\cal C}\; n (1-\alpha^2) T^{3/2}
\end{equation}
The typical relaxation time is then  
$\tau_{\text{relax}}\sim [{D}/({n(1-\alpha^2)})]^{2/3}$. In the small
inelasticity limit $(\eps\rightarrow 0)$, this relaxation time
diverges as $\tau_{\text{relax}}\sim \eps^{-2/3}$. We have indeed
observed such a behaviour of the approach to the steady state both in MD 
and DSMC simulations.

Many pieces of information on the stationary distribution function have been
obtained in \cite{vNE}; in particular, the deviations from a Gaussian
$\Phi(c)=\ee^{-c^2/2}/\sqrt{2\pi}$
have been investigated by the Sonine expansion~\cite{Landau}
\begin{equation}
f(c) = \Phi(c) \left( 1 + \sum_{p=1}^{\infty} a_p S_p (c^2)\right) \ ,
\label{eq:sonine}
\end{equation}
where the $S_p$'s are polynomials orthogonal with the Gaussian weight $\Phi$.
The coefficients $a_p$ are then obtained from the moments of $f$.
From the definition of temperature,
$a_1$ vanishes and the first correction $a_2$, which is related to the kurtosis 
of the velocity distribution, has been computed in any dimension
neglecting non-linear contributions of ${\cal O}\left(a_2^2\right)$; 
in dimension $1$, it has the expression
\begin{equation}
a_2 \equiv \frac{4\, \langle v^4\rangle}{3\, \langle v^2 \rangle^2} -1 
= \frac{16 (1 - 2 \alpha^2)}{129 + 30 \alpha^2}  \ .
\label{eq:a2}
\end{equation}
We note a peculiarity of dimension $1$: $a_2$ does not 
vanish as $\alpha\to 1$, unlike in space dimensions $d>1$ in which 
$\lim_{\alpha \to 1} a_2=0$. This is a hint that the quasi-elastic
limit is more singular in $d=1$ than in higher dimensions. 

Besides it was shown in \cite{vNE} how to determine the high energy tail 
of the velocity
distribution. We briefly recall the argument. The $\eps$-dependent gain 
term in the collision integral appearing in the
Boltzmann equation~(\ref{ludwig}) is {\it a priori} neglected. In the large 
velocity limit the resulting
equation for the steady-state distribution $P_s(v)$ reads
\begin{equation}
D\frac{\dd^2 P_s}{\dd v^2}=-n|v|P_s(v)
\end{equation} 
which yields a high energy tail of the form
$\exp(-\frac{2}{3}\sqrt{n/D}|v|^{3/2})$. Then one verifies that the gain
term is indeed negligible (as would be the case for any solution decaying 
faster than exponentially).
The $3/2$ exponent is independent of the space dimension; therefore
the behaviour of the large $c$ tail is singular for $\eps \to 0$,
with an exponent jumping from $3/2$ for $\eps > 0$ to $2$ for $\eps=0$
(for dimensions larger than $1$, the elastic system equilibrates and thus
$f$ is a Gaussian).

\subsubsection{Numerical solution of the homogeneous Boltzmann equation}

The DSMC method gives access numerically to the exact solution of the 
Boltzmann equation, and we restricted our study 
to the homogeneous situation.
We have obtained the velocity distributions
for various values of the restitution coefficient. Another powerful iterative
method was recently introduced by Biben {\it et al.}~\cite{Biben}.
Let us recall the idea of this method:

The stationary velocity distribution can be obtained numerically through a direct
iteration of equation (\ref{ludwig}). From an initial guess for the velocity 
distribution
\( P(v) \) (a step function for example) the time evolution can be computed
from equation (\ref{ludwig}) until the steady state is reached. 
Taking advantage of the
\( v\to -v \) symmetry of the velocity distribution, we only need to know the
values of \( P \) for positive velocities. \( P(v) \) is then discretized
from \( v=0 \) to \( v=(N_{v}-1)dv \) where typically 
\( N_{v}=1000 \) and \( dv=0.01\sqrt{D\sqrt{\pi }/(n(1-\alpha ^{2}))} \). 
The right hand side of equation (\ref{ludwig}) can be estimated 
using Simpson integration,
combined with a quadratic interpolation method to estimate the values of 
$$ P\left( \frac{2v-(1-\alpha )v^\prime }{1+\alpha }\right)  $$
whose argument do not necessarily coincide with the velocity
discretization. An implicit method is
used to solve the diffusion equation: if \( P_{i}^{t+dt} \) denotes the new
estimate of \( P(i\: dv) \) at time \( t+dt \), the left hand side of equation
(\ref{ludwig}) can be written in the time-velocity discretized space:
\[
\frac{dP}{dt}-D\partial^2 _{v}P\to \frac{P_{i}^{t+dt}-P_{i}^{t}}{dt}-D\frac{P_{i+1}^{t+dt}-2P_{i}^{t+dt}+P_{i-1}^{t+dt}}{dv^{2}}\]
which leads to the following equation for \( P_{i}^{t+dt} \):
\begin{equation}
\left( 1+2D\frac{dt}{dv^{2}}\right) P_{i}^{t+dt}-D\frac{dt}{dv^{2}}
\left( P_{i+1}^{t+dt}+P_{i-1}^{t+dt}\right)=P_{i}^{t}+
dt\left\{ \hbox {RHS of (\ref{ludwig}) at site $i$ and time $t$}\right\}.
\end{equation}
We recognize a band tridiagonal matrix in the left hand side which can easily
be inverted numerically to provide the new value of the velocity distribution
at time \( t+dt \) . Normalization of the velocity distribution is enforced at
each time step.

\begin{figure}[ht]
\centerline{
\epsfig{figure=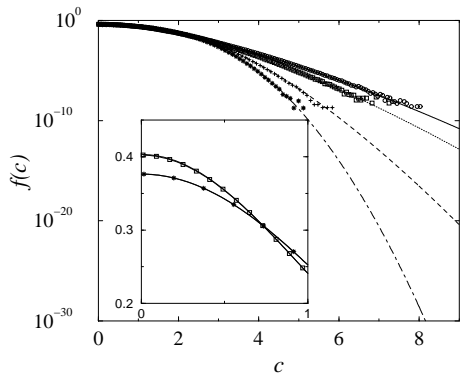,width=6cm,angle=0}    }
\vskip 2mm       
\caption{Velocity distributions obtained by DSMC with 25000 particles 
(symbols) or by the iteration method (lines), in a log-linear scale,
for restitution coefficients $0.1$, $0.6$, $0.9$ and $0.99$ 
(from top to bottom). Inset: same distributions on a linear scale,
for restitutions 0.6 (stars) and 0.99 (squares).}
\label{fig:dsmc_biben}
\end{figure}

Figure 
\ref{fig:dsmc_biben} shows a perfect agreement between the two methods,
the iterative method allowing to reach a much higher
precision (see the $y$-scale).
The obtained distributions show strong deviations from the Gaussian, just as 
in the MD case.
However, the distributions obtained by MD and DSMC agree only in the
$\alpha \to 1$ limit, as was expected because of the spatial
inhomogeneities appearing in the MD simulations.

\begin{figure}[ht]
\centerline{
\epsfig{figure=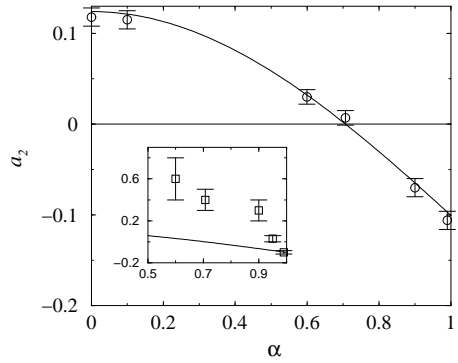,width=6cm,angle=0}    }
\vskip 2mm        
\caption{Values of the second Sonine coefficient $a_2$, obtained by
measuring the fourth cumulant of $f(c)$ in DSMC simulations (circles),
together with the kinetic theory prediction Eq. \ref{eq:a2} (line). Inset: MD
values for $a_2$ (squares), and kinetic theory prediction (line).
}
\label{fig:a2}
\end{figure}

The measure of the fourth cumulant $a_2$ (the first correction to the Gaussian)
shows an excellent agreement between the DSMC
data and the kinetic theory predictions from equation (\ref{eq:a2}), 
on the whole range of inelasticities (see Fig. \ref{fig:a2}).
In the limit $\alpha \to 1$, $a_2$ obtained in MD coincides with the 
prediction of Eq. (\ref{eq:a2}), namely, $a_2 \to -16/159$. Moreover, the
full velocity distribution function coincides with that obtained in DSMC
(see Fig. \ref{fig:md_dsmc.999} below). However, as $\alpha$ decreases, 
the MD results significantly deviate from their molecular chaos counterpart
(see the inset of Fig. \ref{fig:a2})

\begin{figure}[ht]
\centerline{
\epsfig{figure=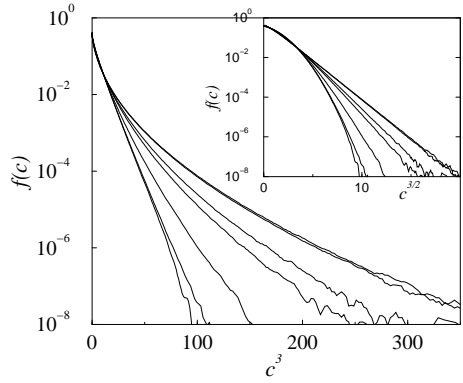,width=6cm,angle=0}    }
\vskip 2mm       
\caption{Velocity distributions obtained by DSMC with 25000 particles, versus
$c^3$ and in the inset versus $c^{3/2}$, for restitution coefficients 
$0$, $0.1$, $0.6$, $0.707$, $0.9$, $0.99$, $0.999$ (from top
to bottom): at low restitution coefficient $f(c)$ shows an
$\exp( - A c^{3/2})$ behaviour, and an $\exp( - A c^3)$ behaviour as
$\alpha$ goes to $1$.}
\label{fig:vcube}
\end{figure}

Moreover, plotting the velocity distribution versus either $c^{3/2}$ or
$c^3$, as in figure \ref{fig:vcube}, shows that the high energy
tail obtained by the DSMC method has a shape going from
the predicted $\exp( - A c^{3/2})$ at large inelasticity
to an $\exp( - A c^3)$ behaviour at low inelasticity. For intermediate
values of $\alpha$, a fit to a form $\exp( - A c^B)$ would lead to
intermediate values of $B$. Since the {\em real} high energy limit
{\em has to} follow an $\exp( - A c^{3/2})$ law \cite{vNE}, this shows 
that for low $\eps$, this limit
is far beyond reach of usual numerical methods, and emphasizes the fact that
the range over which the large $c$ limit is valid depends on the
inelasticity. This is an important point since
experiments have a limited precision, and the distribution function
will have a practically vanishing weight much
{\em before} this range is reached.  
In the next subsection we will see that the investigation of the
$\alpha \to 1$ limit and the use of the iterative method allows
to understand the $\exp( - A c^3)$ form obtained for $\alpha$
close to $1$.

\subsection{Small inelasticity limit} 
\begin{figure}[ht]
\centerline{ 
\epsfig{figure=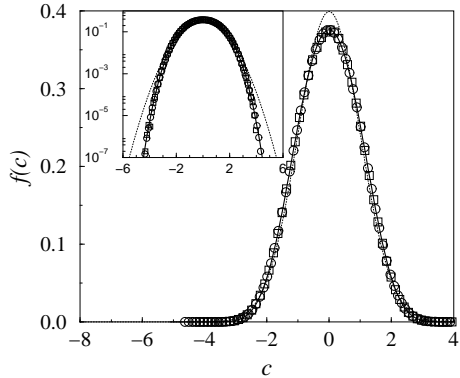,width=6cm,angle=0}    }
\vskip 2mm       
\caption{Symbols: velocity distributions obtained numerically
at $\alpha = 0.999$ by MD and DSMC simulations; the solid line
is the numerical solution $\phi(x)$ of Eq. (\ref{inteq}), corresponding to the
$\alpha \to 1$ limit, and here rescaled in order to have the same
variance as the simulation data ($\langle c^2 \rangle=1$). 
The dotted line is the Gaussian distribution.}
\label{fig:md_dsmc.999}
\end{figure}

For small values of $\eps\equiv 1-\alpha$, the Boltzmann equation takes the form
\begin{equation}
\p_t P(v,t)=D\p^2_vP(v,t)
+n\eps\int\limits_{-\infty}^\infty\dd
w|v-w|P(w,t)\left[P(v,t)+\frac{1}{2}(v-w)\p_vP(v,t)\right]+{\cal O}(\eps^{2})
\label{Boltzmanneps}
\end{equation}
The $\eps\to 0$ limit can now be taken by introducing a scaled velocity
$x=({n}\eps/{D})^{1/3}v$. Using $x$, the Boltzmann equation yields
an equation for the scaled distribution function
$\phi(x)=(n\eps/ D)^{1/3}P_s(v)$ (where $P_s(v)$ is the stationary
distribution $\lim_{t \to \infty}P(v,t)$):
\begin{equation}
\frac{\dd^2\phi}{\dd x^2}+\int\limits_{-\infty}^\infty\dd y|x-y|\phi(y)\left[
\phi(x)+\frac{1}{2}(x-y)\frac{\dd\phi}{\dd x}\right]=0 \,
\end{equation}
and we can integrate this equation twice to obtain 
\begin{equation}
\phi(x)=C\,
\exp\left\{-\frac{1}{6}\int\limits_{-\infty}^\infty\dd y|y-x|^3\phi(y)\right\} \, .
\label{inteq}
\end{equation}
Here $C$ is a constant determined from the normalization 
condition $\int \dd x \phi(x)=1$. We have used the above equation to 
implement an iterative scheme to find numerically the corresponding
velocity distribution. Moreover, as already pointed out in
\cite{Benedetto}, one can easily see that 
the large $|x|$ limit is given by: 
\begin{equation}
\phi(x)=C\ee^{-\frac{1}{6} |x|^3} \, ,
\label{largexasymp}
\end{equation}
while at small $x$ the function can be approximated by a Gaussian
\begin{equation}
\phi(x)={\widetilde C}\,\ee^{-\frac{1}{2}\lambda x^2}
\label{smallxasymp}
\end{equation}
with $\lambda=\int\dd x\;|x|\phi(x)\approx 0.785$ determined from 
the numerical 
solution of Eq. (\ref{inteq}) and $\tilde C=C\exp(-\int\dd x|x|^3\phi(x)/6)$.
The full numerical solution (displayed in Fig. \ref{fig:md_dsmc.999})
can also be investigated for locating the place where the crossover 
between the Gaussian and the $\exp{(-|x|^3/6)}$ type behaviours takes place. 
We find that the crossover range deduced from comparing the asymptotics 
(\ref{largexasymp}) and (\ref{smallxasymp}),
$x_{\text{cr}}=3\lambda\simeq 2.36$, is actually in agreement with
numerical observations of the full function. Thus returning to 
non-scaled velocities, we can see that the crossover velocity 
$v_{\text{cr}}^{(1)}$ diverges in the $\eps\rightarrow 0$ limit as 
\begin{equation}
v_{\text{cr}}^{(1)}\approx \left(\frac{D}{n\eps}\right)^{1/3} \, .
\end{equation}
The important consequence of this result is that, for small
dissipation, the $\exp (-A v^3)$ regime is reached for
larger and larger velocities. 

\begin{figure}[ht]
\centerline{ 
\epsfig{figure=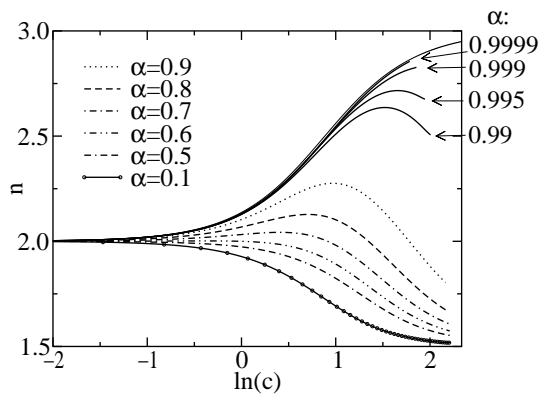,width=6cm,angle=-90}    }
\vskip 2mm       
\caption{Effective exponent defined by Eq. (\ref{eq:exp}), for the solution of
the Boltzmann equation obtained by the iterative method, and for the 
numerical solution of the $\alpha \to 1$ limit [upper dotted curve, obtained 
from the iterative resolution of Eq. (\ref{inteq})].
}
\label{fig:exponent}
\end{figure}

At large but finite velocities, an effective exponent can be defined 
by
\begin{equation}
n= \frac{\dd}{\dd \ln v} \ln \left\{ -\ln \left[\frac{P(v)}{P(0)}\right] \right\} \ ,
\label{eq:exp}
\end{equation}
corresponding to an apparent $\exp(-A v^n)$ behaviour.
The values of $n$ for various inelasticities, obtained with the
iterative method, are displayed
in figure \ref{fig:exponent}, together with the $\alpha \to 1$ limit.
The effective exponent, starting from $2$ at small velocities, increases
and reaches a maximum at velocities that scale as 
$\eps^{-1/3}$ for small $\eps$ (as $v_{\text{cr}}^{(1)}$). 
The height of the maximum, $n_{\text{cr}}^{(1)}$ scales as
$3 - n_{\text{cr}}^{(1)} \sim \eps^{1/3}$. 

One should note that the above scaling analysis explores only the
$v\sim \eps^{-1/3}$ range of the velocity space. As already mentioned, 
for any fixed $\eps$, the large $v$ limit 
of the distribution function is a stretched exponential
\begin{equation} 
P_s(v)\sim \exp\left\{-\frac{2}{3}\left(\frac{n}{D}\right)^{1/2}|v|^{3/2}\right\} \, ,
\label{verylargev}
\end{equation}
which is indeed observed numerically at relatively low values
of $\alpha$ (see figure \ref{fig:vcube}).
Comparing the arguments of the exponents in (\ref{largexasymp}) and 
(\ref{verylargev}) one finds a crossover scale diverging as 
\begin{equation}
v_{\text{cr}}^{(2)}\approx \left(\frac{D}{n\eps^2}\right)^{1/3} \, .
\end{equation}

The effective exponent displayed in figure \ref{fig:exponent} indeed decreases
at velocities larger than $v_{\text{cr}}^{(1)}$. For large inelasticities,
the $n \to 3/2$ limit of large velocities is observed; however,
since $v_{\text{cr}}^{(2)}\approx v_{\text{cr}}^{(1)}/\eps^{1/3}$, it 
becomes impossible to observe the asymptotics (\ref{verylargev}) 
for small inelasticities, even for almost realistic values of $\alpha$ like 
$0.95$ (in experiments, $\alpha \sim 0.8\ -\ 0.9$).

It has to be emphasized that this kind of behaviour is also observed
in higher dimensions (for example, simulations of a two-dimensional
heated granular gas with $\alpha =0.8$ yield an almost Gaussian 
velocity distribution, even if the predicted high energy behaviour
is $\exp( - v^{3/2})$); it is therefore to be kept in mind for
the comparisons of models with experiments in which the available precision
often does not allow to reach the predicted tails.

Finally, as already mentioned, it can be seen in Fig. \ref{fig:md_dsmc.999} that
for $\eps \to 0$, there is an excellent agreement 
between the single particle velocity distribution obtained in
MD simulations (including both the space and velocity correlations),
and that derived either from the asymptotic ($\eps\to 0$) 
distribution function $\phi(x)$
or from the Monte Carlo simulation of the Boltzmann equation.

\section{Freely cooling system}
\label{sec:free}

\subsection{General considerations}
\label{ssec:general}

In the freely cooling system, no energy is injected and the temperature
is monotonically decreasing with time. 
The first investigations of the one-dimensional freely evolving
gas were undertaken in \cite{Mcnamara1,Mcnamara2,Goldhirsch,Sela};
it was shown by Molecular  Dynamics simulations that, depending on the values 
of the number of particles
and of the restitution coefficient, different instabilities could
occur: e.g. at fixed number of particles
$N$, if $\alpha$ is lower than a threshold, strong clustering
occurs and leads to inelastic collapse \cite{Mcnamara1}.
At larger $\alpha$, the instability develops more slowly, and
the inelastic collapse is avoided. The temperature then decays according to
the rate equation $dT/dt \propto -T^{3/2}$, i.e. $T(t) \sim t^{-2}$ \cite{haff},
however derived for an homogeneous system, whereas
strong heterogeneities develop both in velocity and space
coordinates; a wavy, oscillatory in time, perturbation appears in a 
``phase-space'' plot (velocity versus position) \cite{Mcnamara2,Goldhirsch},
with a tendency to form a bimodal velocity distribution.

The choice of a suitable
quasi-elastic limit (where $\eps \to 0$ and $N \propto 1/\eps$
to avoid the inelastic collapse) leads to
a simplified Boltzmann equation
\cite{Mcnamara2,Sela,Du,Ramirez}, which can be understood
using arguments from kinetic theory and hydrodynamics.
This equation can be
considered as formally exact as it has also been derived in \cite{Caglioti}
from the exact
BBGKY-like hierarchy governing the evolution of the distribution 
\cite{Resibois}.
In this quasi-elastic limit,
the velocity was observed to develop a two-hump structure
reminiscent of the bimodal velocity distribution
observed in Molecular Dynamics \cite{Mcnamara2,Sela}.
Moreover, exact results were derived in the context 
of the above-mentioned limit, where it was shown 
in particular that to leading order in $\eps$, 
the velocity distribution consists
of two symmetric Dirac peaks \cite{Caglioti}.

\begin{figure}[ht]
\centerline{ 
\epsfig{figure=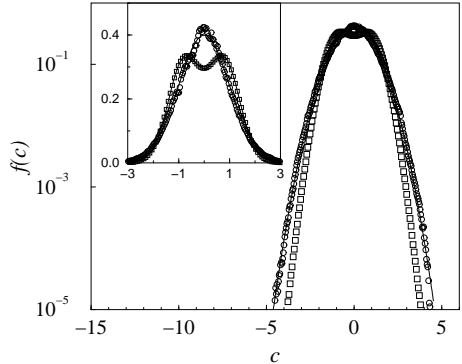,width=6cm,angle=0}    }
\vskip 2mm       
\caption{Rescaled velocity distributions at large times at $\alpha=0.85$
obtained by MD with $N=25000$ (circles) and DSMC (squares) simulations. 
The solid line is the Gaussian distribution. In MD simulations, the 
inelastic collapse
has been regularized by considering the same modified collision rule as in
{\protect\cite{bennaim}}.
}
\label{fig:freea.85}
\end{figure}

Extensive MD simulations were carried out in \cite{bennaim}, using
large sizes and probing large times, starting from an homogeneous situation 
with an initial given velocity distribution.
As long as the system is homogeneous, the temperature decays according to
$T(t) \sim t^{-2}$ \cite{haff}.
As time evolves, space clustering of particles occurs 
and a $t^{-2/3}$ decay is obtained \cite{bennaim}. Since the number of
particles is large, inelastic collapse should then occur; this catastrophic 
event is avoided by imposing that
the collisions between particles with relative velocity
smaller than a given threshold are elastic (they can equally be chosen
sticky), and it was checked that 
the results do not depend on the chosen threshold. 
In this respect, the authors of \cite{bennaim}
showed that the one-dimensional inelastic fluid
belongs to 
the ``universality class'' of the sticky gas, and advocated a mapping to the
Burgers equation to describe the appearing heterogeneities.
Moreover, at large times the rescaled stationary velocity
distribution $f(c)$ was found very close to a Gaussian up to the available
accuracy (see also Figure \ref{fig:freea.85}),
even if the mapping to the Burgers equation predicts
an $\exp (- A c^3)$ high energy tail. 
In fact, the bimodal structure of $f(c)$ reported in \cite{Mcnamara2,Sela} can also
be observed in this case during the
transient homogeneous behaviour, during which spatial 
heterogeneities and correlations
build up, as shown in \cite{Baldassarri}; moreover, it can
be clearly seen only by a convenient choice of the initial 
velocity distribution.
The importance of spatial heteregeneities and correlations is emphasized in 
Fig. \ref{fig:freea.85}, where the velocity distribution obtained 
following the prescription put forward by Ben-Naim {\it et al.}\/ \cite{bennaim}
(that is essentially Gaussian)  is compared to that obtained from the
homogeneous Boltzmann equation \cite{rque}. 
The two-hump structure displayed by the latter
appears for $\alpha>0.8$ and becomes more and more pronounced 
as $\alpha$ increases. 
In the next subsection, we will investigate 
in detail this structure in the $\eps\to 0$ limit.

\subsection{Small inelasticity limit} 

\begin{figure}[ht]
\centerline{
\epsfig{figure=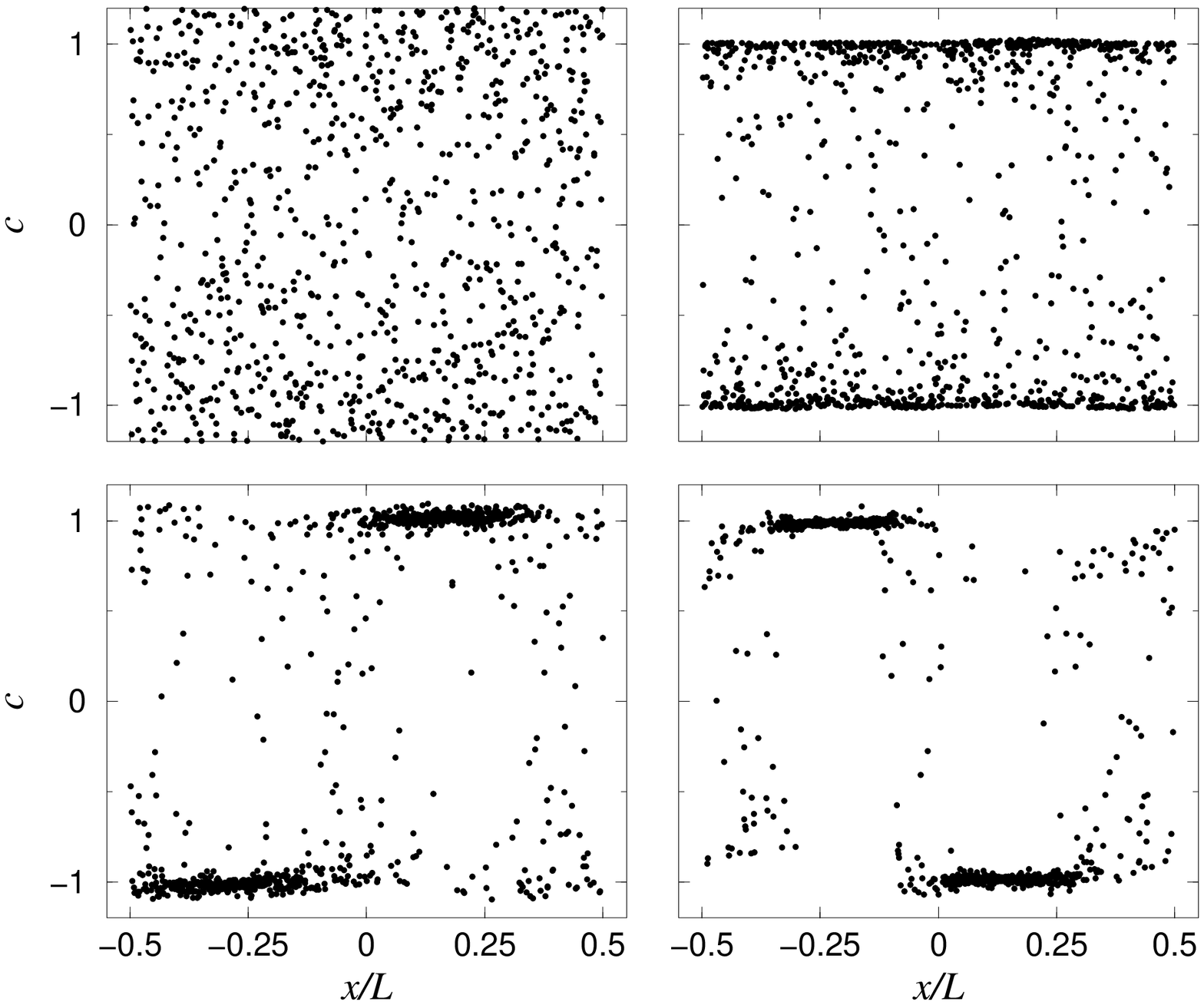,width=6cm,angle=0}    
\hspace{10mm}
\epsfig{figure=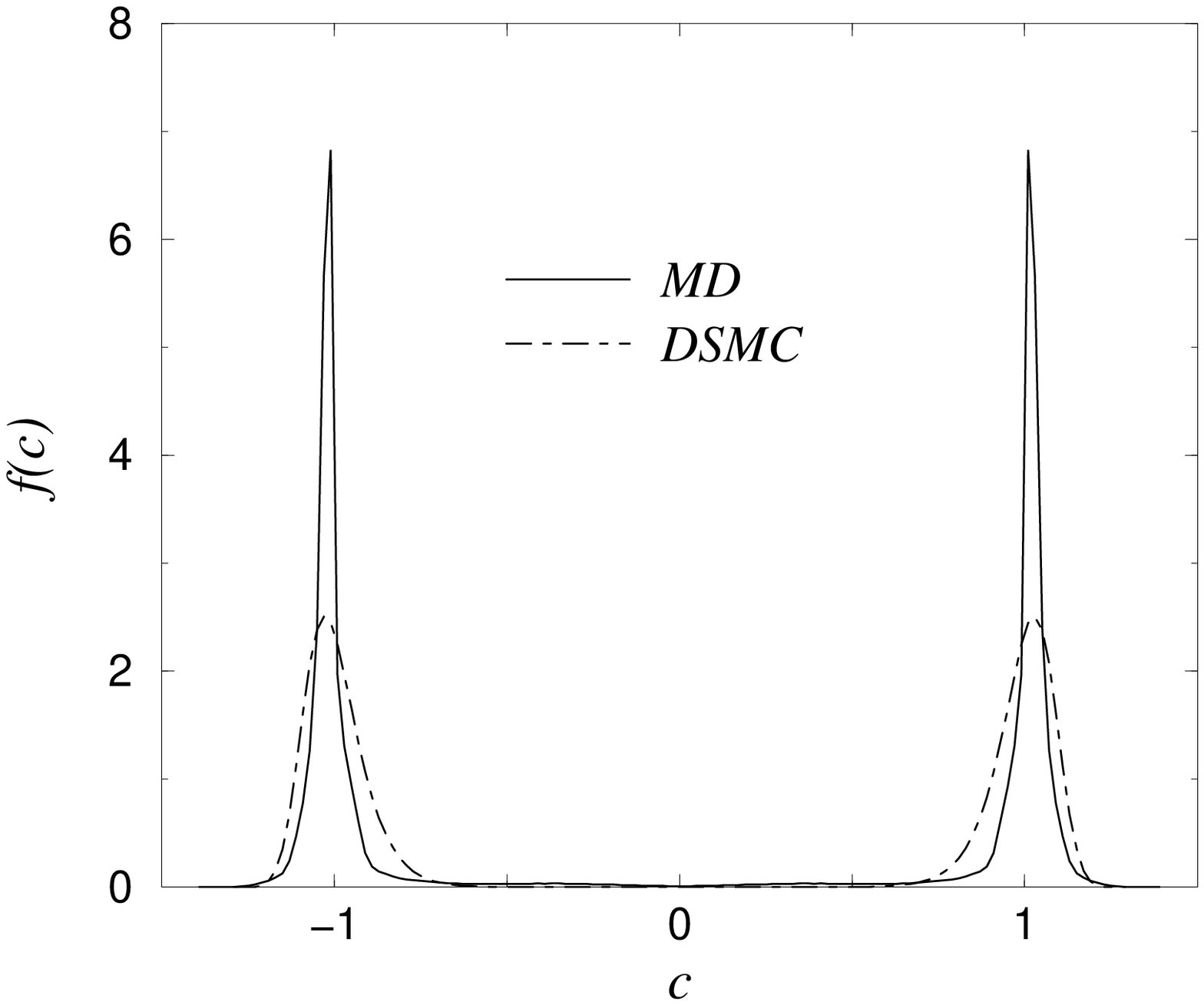,width=6cm,angle=0}    }
\vskip 2mm
\caption{{\em Left:} Velocity-density scatter plots obtained in MD
simulations with $N=1000$ and $\alpha = 0.9995$. Each dot marks the
location of a particle in the $x-c$ plane, where $x$ denotes the 
position in the simulation box and $c$ denotes the velocity
rescaled by the thermal velocity.
Starting from an initial Gaussian distribution of velocities and random 
initial positions, the four snapshots correspond, from left to right and
top to bottom, to four instantaneous configurations observed after
respectively $4.10^3$, $2.10^4$, $4.10^4$ and $6.10^4$ collisions per particle.\\
{\em Right:} Velocity statistics obtained in MD by averaging
in the late time regime for the same initial situation and parameters 
as above, compared to its
``mean-field'' counterpart provided by DSMC simulations at the same inelasticity
($\alpha = 0.9995$). 
}
\label{fig:a9995}
\end{figure}

We have performed MD simulations using the two possibilities to avoid
inelastic collapse mentioned in section \ref{ssec:general}. 
\begin{itemize}
\item Figure \ref{fig:a9995}, left panel displays the results obtained for
$\eps = 5.10^{-4}$, at various stages
of the evolution: at first the system remains homogeneous
but the tendency to form a bimodal velocity distribution is rapidly
obtained. The large time situation consists of
two well defined clusters evolving with opposite
velocities and yielding a sharply
peaked bimodal velocity distribution (Figure \ref{fig:a9995}, right panel).
In this case, the overall kinetic energy 
$E(t) \sim t^{-2}$ decay consists 
in a series of plateaus since most of the dissipation occurs
when the two clusters collide \cite{Mcnamara2}.

\item On the other hand, with the regularization proposed 
in \cite{bennaim}, the duration
of the transient behaviour increases with $\alpha$, but the large time
behaviour of the velocity distribution is always Gaussian.
\end{itemize}

\begin{figure}[ht]
\centerline{
\epsfig{figure=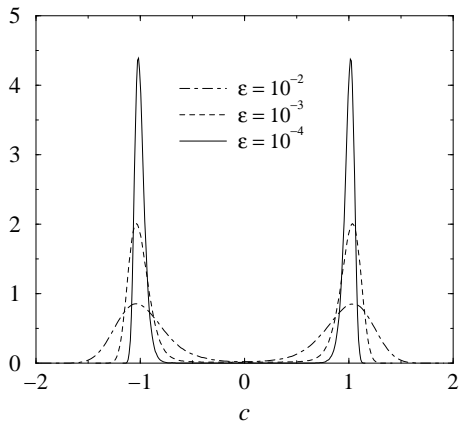,width=6cm,angle=0}    }
\vskip 2mm
\caption{Rescaled velocity distributions at $\alpha=1-\eps$ with 
$\eps=10^{-2}$, $10^{-3}$ and $10^{-4}$, obtained by DSMC simulations.}
\label{fig:freecamel}
\end{figure}

\begin{figure}[htbp]
\centerline{
\epsfig{figure=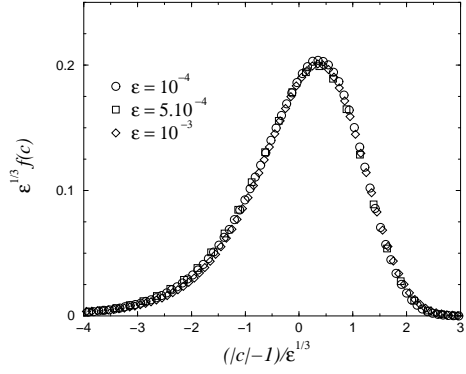,width=6cm,angle=0} }
\vskip 2mm
\caption{Self-similarity of the rescaled distribution functions, 
for small inelasticities.
}
\label{fig:freerescaling}
\end{figure}

It is striking to note that the two above procedures 
(small number of particles or elastic collisions
at small enough velocities) lead to drastically different behaviours
for the decay of the temperature, the spatial
heterogeneities and the velocity distributions.
It is moreover remarkable that the homogeneous solution 
of the Boltzmann equation captures the bimodality of $f(c)$
(see Fig. \ref{fig:a9995}, right panel) associated with 
a strongly heterogeneous system. In order to gain insight 
into the approach to the limit $\eps \to 0$, we therefore 
devote the remainder of this article to the analysis of the scaling
properties of the homogeneous non-linear Boltzmann equation.  
We expect that for low enough inelasticity, $f(c)$ tends towards 
two delta peaks at $c = \pm 1$, as predicted in \cite{Caglioti}. 
Performing DSMC simulations for smaller and smaller
inelasticities allow to approach this regime, and
Figure \ref{fig:freecamel} shows how the
peaks become narrower as $\alpha$ increases.
\noindent As the system cools, the velocity distribution
$P(v,t)$ evolves into the Dirac distribution $\delta(v)$. The above 
numerical results indicate that this distribution actually consists of
two peaks located symmetrically around the velocity origin, at positions
decaying like $\pm (\eps t)^{-1}$. 
Moreover, it appears that the results displayed in Fig. \ref{fig:freecamel} 
for the rescaled velocity $c$ hide a self-similar structure,
with a width of the peaks scaling like $\eps^{1/3}$, as evidenced in Fig. 
\ref{fig:freerescaling} where the distributions corresponding to 
different inelasticities collapse onto a master curve.
The characteristic features of this master curve are investigated below
and to this end, we return to the $\eps$ expansion of the Boltzmann equation.
(\ref{Boltzmanneps}), omitting the Fokker-Planck term $D\partial^2_vP$,
since the fluid evolves freely in the present situation. 
\noindent From the evolution of temperature ($T \propto \eps^{-2} t^{-2}$),
it appears that a convenient scaling variable is $x\equiv n\eps t v$
with a corresponding probability distribution function $\Phi$ related to
$P(v,t)$ through $P(v,t)=n\eps t\Phi(n\eps v t)$. To leading order in $\eps$,
the Boltzmann equation is then
brought into the simple form
\begin{equation}\label{Boltzscaledfreecool}
\frac{\dd(x\Phi)}{\dd x}=\int\dd
y|x-y|\Phi(y)\left[\Phi(x)+\frac{1}{2}(x-y)\frac{\dd \Phi}{\dd x}\right],
\label{eq:eps0}
\end{equation}
the solution of which reads \cite{Caglioti}  
\begin{equation}
\Phi(x)=\frac{1}{2}\left[\delta(x-1)+\delta(x+1)\right].
\end{equation}
Looking for the self-similar structure of the peaks shown in
Figures \ref{fig:freecamel} and \ref{fig:freerescaling} requires
to push the $\eps$-expansion one order further compared to 
Eq. (\ref{eq:eps0})
\begin{equation}
\frac{\dd (x\Phi)}{\dd x}=\int\dd
x'|x-x'|\Phi(x')\left[\frac{1}{(1-\eps/2)^2}\Phi
\left(x+\frac{\eps}{2-\eps}(x-x')\right)-\Phi(x)\right],
\label{eq:eps1}
\end{equation}
and to consider solutions for $\Phi$ of the form
\begin{equation}
\Phi(x)=\eps^{-\nu}\left\{\frac 12
\psi\left(\frac{1+b(\eps)x}{\eps^{\nu}}\right)+\frac 12
\psi\left(\frac{1-b(\eps)x}{\eps^{\nu}}\right)\right\}.
\end{equation}
In Eq. (\ref{eq:eps1}), the positive function $\psi$
has the interpretation of the ($\eps$-rescaled) velocity 
distribution of left (or right) movers,
and $b(\eps)=b_0+\eps^\nu b_1+\eps^{2\nu}b_2$. 
Substituting the scaling assumption for $\Phi$ into (\ref{eq:eps1})
and performing the change of variables $x=-1+\eps^\nu y$ we
obtain an non-linear intro-differential equation for $\psi(y)$:
\begin{eqnarray}
a(\eps)\left[\eps^{1-\nu}(y 
\psi)'-\eps^{1-2\nu}\psi')\right]=
\frac{\eps}{2}\int\dd
y'|y-y'|\psi(y')(\psi(y)+\frac{1}{2}(y-y')\psi'(y))
\nonumber \\
+\frac{\eps^{-\nu}}{2}\int\dd
y'|2-\eps^{\nu}(y+y')|\psi(y')
\times\left[-\eps^{1-\nu}\psi'+\frac{\eps
}{2}(y+y')\psi'+\eps\psi+\frac{1}{2}\eps^{2-2\nu}\psi''\right],
\label{beurk}
\end{eqnarray}
where terms of the form $\psi(-2\eps^{-\nu}+y)$ have been neglected,
anticipating that they will be exponentially small. Terms of order
$\eps^{2-2\nu}, \eps^{1+\nu}$ were equally neglected. Assuming again that $\psi$ 
will have a sharp decay, we write that under the integrals
$|2-\eps^{\nu}(y+y')|=2-\eps^{\nu}(y+y')$. Identifying on both sides of
Eq.~(\ref{beurk}) terms of order $\eps^{1-2\nu}$ and $\eps^{1-\nu}$ leads
respectively  to
\begin{equation}
b_0=\int\dd y'\psi(y')
\end{equation}
which is the normalization condition, and
\begin{equation}
\psi'(y)\left\{b_1+\int\dd y'\;y'\psi(y')\right\}=0
\end{equation}
which relates $b_1$ to $\langle y\rangle$ (a
constant function $\psi$ cannot be a solution). We choose to impose 
$b_1=-\langle y \rangle_\psi=0$, where the notation $\langle ...\rangle_\psi$ 
stands for an average with weight function $\psi$. Then one notices that the
expansion is consistent only to the condition that the $\eps^{2-3\nu}$ term
can be cancelled by order $\eps$ terms, which imposes that
\begin{equation}
\nu=\frac{1}{3},
\end{equation}
and we recover the exponent $1/3$ that was needed to collapse the 
velocity distributions at several small inelasticities, as done empirically
in Fig. \ref{fig:freerescaling}. 
Finally we equate to 0 the terms of order $\eps$ to obtain the following 
integro-differential
equation
\begin{equation}
b_2\psi'(y)
=\int\dd
y'\psi(y')\Bigg[\frac 12\psi(y)\left(|y-y'|-(y+y')\right)
+\frac{1}{4}\psi'(y)\left(|y-y'|(y-y')-(y+y')^2\right)
+\frac 12\psi''(y)\Bigg]
\end{equation}
which we integrate once, remembering that $b_0 = \langle1\rangle_\psi$:
\begin{equation}
b_0\psi'(y)+2b_2\psi(y)+\frac{1}{2}\psi(y)\int\dd 
y'\psi(y')\left(|y-y'|(y-y')-(y+y')^2\right)=0.
\label{eq:psi}
\end{equation}
We know from the direct analysis of the
$\eps\to 0$ limit of the scaling function that $b_0=1$, 
which we use here on. At this stage we note that
integrating the above equation and using that $\langle y\rangle_\psi =0$ leads to
\begin{equation}
2b_2=\langle y^2\rangle_\psi.
\end{equation}
Conversely, setting $b_2=\frac 12 \langle y^2\rangle_\psi$ will 
automatically enforce $\langle y\rangle_\psi =0$. 
We rewrite the equation for $\psi$ in the form
\begin{equation}
\ln\psi(y)=\ln C-2b_2 y+\frac{1}{6}\int\dd 
y'\psi(y')\left[|y-y'|^3-(y+y')^3\right].
\end{equation}
We now investigate the asymptotics of $\psi$. The left
tail of the distribution at large negative values of $y$ reads:
\begin{equation}
y\to-\infty,\;\;\psi(y)\simeq C\exp\left[\frac{1}{3}y^3+o(1)\right].
\end{equation}
This sharp decay at $y\to-\infty$ {\it a posteriori} justifies the
approximations made in the course of the calculation. 
Note that omitting the first line in the rhs of Eq.~(\ref{beurk}) leads to 
exactly the same
behaviour of the tail of the distribution. This has a physical
interpretation: collisions between particles heading in the same direction can 
be neglected at large velocities.
Opposite-velocity collisions are responsible for the form of the tail at large
velocities. The $o(1)$ represents contributions going exponentially
fast to 0.

The right tail $y\to+\infty$ decays exponentially fast as
\begin{equation}
\psi(y)\,\simeq \,C'\,\exp\left\{-2\langle y^2\rangle_\psi \; y+o(1)\right\}\quad
\hbox{with}\quad
C'\,=\,C\,\exp\left\{-\frac{1}{3}\langle y^3\rangle_\psi\right\}
\end{equation}
which again justifies the assumptions made so far. The $o(1)$ again represents 
contributions going exponentially
fast to 0. 
For completeness we mention the $y\to 0$ behaviour of the scaling function:
\begin{equation}
\psi(y)=C''\exp\left\{-\frac{y}{2}\langle 3s^2+|s|s
\rangle_\psi+y^2\langle|s|\rangle_\psi)+{\cal
O}(y^3)\right\}, \quad \hbox{with} \quad 
C''=C\exp\left\{\frac{1}{6}\langle|y|^3-y^3\rangle_\psi\right\}.
\end{equation}

\begin{figure}[ht]
\centerline{
\epsfig{figure=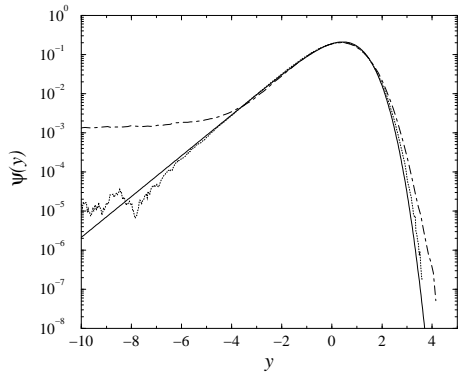,width=6cm,angle=0} }
\vskip 2mm
\caption{Comparison of the scaling function $\psi(y)$ (see text for definition)
obtained within DSMC ($\eps = 10^{-4}$, dotted curve and
$\eps=10^{-3}$, dashed curve), with the solution of Eq. (\ref{eq:psi})
corresponding to the quasi-elastic limit. }
\label{fig:freerescaling_log}
\end{figure}

As a side remark, the integro-differential
equation for $\psi$ can be cast in the form of an ordinary 
fourth-order differential equation
\begin{equation}
(\ln\psi)^{(\text{iv})}=\psi
\end{equation}
Finally, we note that numerical iteration of the integro-differential equation
(\ref{eq:psi}) converges extremely rapidly. This allows to determine 
the numerical constants $C$, $\langle y^2\rangle_\psi$, 
$\langle y^3\rangle_\psi$ appearing in the asymptotics.
The results obtained from this numerical scheme are compared with those
of the DSMC method in Fig. \ref{fig:freerescaling_log}, and show a quantitative
agreement which improves as $\eps$ is lowered in DSMC, as expected
(see the dotted curve at $\eps=10^{-4}$, closer to the asymptotic
scaling form for $\psi$ than the dashed line corresponding to $\eps=10^{-3}$).

\section{Conclusion}

We have investigated the velocity statistics of 
one-dimensional granular fluids of
inelastic particles, with a particular emphasis on scaling properties
in the elastic limit,
both in the absence of an external forcing and in a system heated by random 
accelerations. For the heated system, we showed that the expected high energy
tail $\sim \exp(-A c^{3/2})$ yields the correct asymptotic behaviour at finite 
inelasticity $\eps$, but this asymptotics
is masked by a tail $\sim \exp(-B c^{3})$ for 
$\eps \to 0$, with the {\em rescaled} crossover velocity between the 
two regimes scaling like 
$\eps^{-1/3}$. This shows that the limits of high velocity and low inelasticity 
do not commute: if $\eps \to 0$ is taken {\em before} the high energy limit,
the distribution exhibits an asymptotic $\sim \exp(-A c^{3})$ large $c$ behaviour:
\begin{eqnarray}
&&f(\eps,c) \,\, \stackrel{c\to\infty}{\propto} \,\,
\exp(-A \,c^{3/2}) \quad \hbox{for any $\eps \ne 0$} 
\label{eq:tail1}\\
\hbox{whereas~~} 
\lim_{\eps\to 0}\,\,&& f(\eps,c) \,\, \stackrel{c\to\infty}{\propto} \,\, \exp(-A\, c^{3}). 
\label{eq:tail2}
\end{eqnarray}
Thanks to a high precision iterative scheme allowing to solve the homogeneous 
non-linear Boltzmann equation, we could obtain the velocity distribution over
30 orders of magnitude at arbitrary $\eps$, and thus define the apparent
exponent $n$ of the stretched exponential law for large $c$ 
$[f(c)\propto  \exp(-Cc^n)]$. However, even with such a precision, 
the crossover between the two behaviours (\ref{eq:tail1}) or
(\ref{eq:tail2}) is difficult to investigate
(see Fig. \ref{fig:exponent}).

For the freely evolving 1D granular fluid, we have investigated in detail
the structure and scaling properties of the two-hump
velocity distribution exhibited by the solution
of the homogeneous cooling state of the Boltzmann equation, both numerically
and analytically. Such a bimodal distribution captures an
essential feature of the velocity distributions obtained in Molecular Dynamics
simulations for parameters hindering the inelastic collapse ({\it i.e.}
extremely small $\eps$ or small systems). In this respect, a perturbative
Sonine expansion in the spirit of that put forward in \cite{vNE}
fails at small $\eps$, whereas such an expansion 
turned out to be remarkably accurate for 
the heated gas (see Fig. \ref{fig:a2}).
In both cases it would predict a non vanishing
kurtosis correction $a_2$ for $\eps \to 0$, which is a peculiarity of
$d=1$; as soon as $d>1$, $a_2$ vanishes for small inelasticities, 
with or without forcing.

It would be of interest to perform the same analysis for realistic 
space dimensions $d>1$, and to quantify more precisely 
space and velocity correlations \cite{Soto}, 
for both the heated and unforced systems.

\acknowledgements
Z.R. has been partially supported by the Hungarian Academy of
Sciences (Grant No. OTKA T029792).

\end{document}